\magnification=1100

\raggedbottom

\font\gross=cmbx12 scaled \magstep2
\font\mittel=cmbx10 scaled\magstep1

\font\sf=cmss10
\font\pl=cmss10 scaled \magstep1

\nopagenumbers
\headline={\ifnum\pageno=1\hfill\else
{\sf I. G. Avramidi and G. Esposito :
Heat-kernel asymptotics with generalized 
boundary conditions}
\hfill\rm\folio\fi}
\def\II{{\rm 1\!\hskip-1pt I}}

\def\Artanh{{\rm Artanh\,}}

%-----------------------------------

{\null
\vskip-1.5cm
\hskip7truecm{ \hrulefill }
\vskip-.55cm
\hskip7truecm{ \hrulefill }
%\vskip1.5mm 
\smallskip
\vskip1mm
\hskip7truecm{{\pl{} University of Greifswald (January, 1997)}} 
%\smallskip 
\smallskip
%\vskip0.1mm
\hskip7truecm{ \hrulefill }
\vskip-.55cm
\hskip7truecm{ \hrulefill } 
\bigskip

\hskip7truecm{\ hep-th/9701018}
\smallskip
\hskip7truecm{\ Revised on November 1997}
\bigskip
\hskip7truecm{\ Submitted to:}
\smallskip
\hskip7truecm{\ \sf Classical and Quantum Gravity}

\vfill

\centerline {\gross Heat-kernel asymptotics with}
\smallskip
\centerline{\gross generalized boundary conditions}
\bigskip

\centerline {{\mittel Ivan G Avramidi}$^{1,}$
\footnote{$^*$}{On leave of absence from Research Institute
for Physics, Rostov State University, Stachki 194,
344104 Rostov-on-Don, Russia. E-mail: 
avramidi@rz.uni-greifswald.de}
and {\mittel Giampiero Esposito}$^{2,3}$}
\bigskip
\noindent
\centerline
{\it ${ }^{1}$Department of Mathematics, University
of Greifswald,}
\centerline{\it  Jahnstr. 15a, 17487 Greifswald, Germany}
\medskip
\centerline
{\it ${ }^{2}$Istituto Nazionale di Fisica Nucleare,
Sezione di Napoli}
\centerline{\it Mostra d'Oltremare Padiglione 20,
80125 Napoli, Italy}
\medskip
\noindent
\centerline
{\it ${ }^{3}$Universit\`a di Napoli Federico II, 
Dipartimento di Scienze Fisiche}
\centerline{\it Mostra d'Oltremare Padiglione 19,
80125 Napoli, Italy}

\vfill
\noindent
The quantization of gauge fields and gravitation 
on manifolds with boundary makes it
necessary to study boundary conditions which involve both
normal and tangential derivatives of the quantized field.
The resulting one-loop divergences can be
studied by means of the asymptotic expansion of the heat 
kernel, and a particular case of their
general structure is here analyzed in
detail. The interior and boundary contributions to heat-kernel
coefficients are written as linear combinations of all
geometric invariants of the problem. The behaviour of the
differential operator and of the heat kernel under conformal
rescalings of the background metric leads to recurrence 
relations which, jointly with the boundary conditions,
may determine these linear combinations.
Remarkably, they are expressed in terms of universal functions, 
independent of the dimension of the background and invariant
under conformal rescalings, and new geometric invariants
contribute to heat-kernel asymptotics. Such technique is applied
to the evaluation of the $A_{1}$ coefficient when the matrices
occurring in the boundary operator commute with each other.
Under these assumptions, the form of the
$A_{3/ 2}$ and $A_{2}$ coefficients is obtained for 
the first time, and new equations among universal functions
are derived. A generalized formula, relating asymptotic heat
kernels with different boundary conditions, is also obtained.
\vskip .3cm
\leftline{PACS numbers: 0370, 0460}
\eject

\leftline{\mittel 1. Introduction}
\bigskip

In several branches of classical and quantum field theory,
as well as in the current attempts to develop a quantum 
theory of the universe and of gravitational interactions,
it remains very useful to describe physical phenomena in
terms of differential equations for the variables of the
theory, supplemented by boundary conditions for the solutions
of such equations [1, 2]. For example, the problems of
electrostatics, the analysis of waveguides, the theory of
vibrating membranes, the Casimir effect, van der Waals forces,
and the problem of how the universe could evolve from an
initial state, all need a careful assignment of boundary
conditions [1--3]. In the latter case, if one follows a 
path-integral approach [4], one faces two formidable tasks:
\vskip 0.3cm
\noindent
(i) the specification of the geometries occurring in the
``sum over histories" and matching the assigned boundary
data;
\vskip 0.3cm
\noindent
(ii) the choice of boundary conditions on metric perturbations
which may lead to the evaluation of the one-loop semiclassical
approximation.
\vskip 0.3cm
\noindent
Indeed, whilst the full path integral for quantum gravity is a
fascinating idea but remains a formal tool, the one-loop
calculation may be put on solid ground, and appears particularly
interesting because it yields the first quantum corrections to
the underlying classical theory (despite the well known lack
of perturbative renormalizability of quantum gravity based on
Einstein's theory). Within this framework, it is of crucial
importance to evaluate the one-loop divergences of the theory
under consideration. Further to this, the task of the
theoretical physicist is to understand the deeper general
structure of such divergences. For this purpose, one has to
pay attention to all geometric invariants of the problem, in a
way made clear by a branch of mathematics known as invariance
theory [5]. In our paper, however, we shall be concerned with 
a physics-oriented presentation, and hence we here limit 
ourselves to say that the one-loop divergence receives 
contributions from the Riemann curvature of the background,
the extrinsic curvature of the boundary, a gauge curvature
(if a gauge field is studied), jointly with some matrices 
occurring in the differential operator and in the boundary
conditions of the problem. All these ingredients will become
sufficiently clear in the following sections.

The recent investigations in Euclidean 
quantum gravity [2, 6--9] and
quantized gauge fields [10, 11] on manifolds with boundary have
shown that boundary conditions involving both normal and 
tangential derivatives of a field arise naturally if one
requires invariance of the whole set of boundary conditions
under infinitesimal diffeomorphisms [2, 6--9] 
or Becchi-Rouet-Stora-Tyutin (BRST)
transformations [11]. Thus, shortly after the correct formulae
for heat-kernel asymptotics with a mixture of Dirichlet and
Robin boundary conditions were obtained for the first time
[5, 12] after several decades of dedicated efforts by many authors,
it became clear that, even just for Laplace type operators on
manifolds with boundary, the most difficult piece of work in
spectral geometry [5] is yet to be done. We are thus interested
in a framework whose description is as follows.

A smooth manifold $M$ of dimension $m$ is given, with Riemannian (i.e.
positive-definite) metric $g$. Moreover, a vector bundle $V$ 
over $M$ is given with connection $\nabla$
and curvature $\Omega$, say. 
There is also a smooth $(m-1)$-dimensional boundary $\partial M$ with
the induced metric $\gamma$ and the resctriction $\widehat \nabla$ of 
$\nabla$ to $\partial M$. 
An operator of Laplace 
type is a second-order elliptic operator on $M$ which can be
put in the form (summation over repeated indices is
understood)
$$
P=-g^{ab}\nabla_{a}\nabla_{b}-E 
\eqno (1.1)
$$
where $E$ may be viewed as a ``matrix"-valued function. 
In general, one studies tensor or 
spinor fields, for which
the operator $P$ carries a finite number of discrete indices. 
The general framework for the investigation of 
one-loop ultraviolet divergences in quantum field theory, consists
of the heat-kernel approach with the corresponding
$\zeta$-function regularization [2, 13].
The heat kernel is, by definition, a solution of the heat equation
$$
\left({\partial \over \partial t}+P\right)
U(x,x';t)=0 
\eqno (1.2)
$$
for $t>0$, obeying the initial condition
$$
\lim_{t \to 0}\int_{M}{\rm {dvol}}(x')
U(x,x';t)\varphi(x')=\varphi(x) 
\eqno (1.3)
$$
with ${\rm {dvol}}=dx\,\sqrt{\det g}$,
jointly with suitable boundary conditions
$$
[{\cal B}U(x,x';t)]_{\partial M}=0 .
\eqno(1.4)
$$
The functional trace of the heat kernel is obtained by
integrating the trace of the heat-kernel diagonal
$U(x,x;t)$ over $M$, and reads
$$
{\rm {Tr}}_{L^{2}} \; {\rm e}^{-tP}
=\int_{M} 
{\rm {dvol}}(x){\rm {Tr}}\,U(x,x;t)  
\eqno (1.5)
$$
where ${\rm {Tr}}$ is the matrix trace.

In our paper, following [5, 14], we are interested in a slight
generalization of eq. (1.5), where ${\rm e}^{-tP}$ is ``weighted"
with a smooth scalar function $f$ on $M$. 
More precisely, we will be interested in the
asymptotic expansion as $t \rightarrow 0^{+}$ of the functional trace
$$
{\rm {Tr}}_{L^{2}}\Bigr(f{\rm e}^{-tP}\Bigr)
=\int_{M} 
{\rm {dvol}}(x)f(x){\rm {Tr}}\,U(x,x;t) .
$$
The results for the original problem are
eventually recovered by setting $f=1$, but it is crucial to
keep $f$ arbitrary throughout the whole set of calculations,
as will be clear from the following sections.

The asymptotic expansion we
are interested in has the form [5, 10, 14, 15] 
$$
{\rm {Tr}}_{L^{2}}\Bigr(f{\rm e}^{-tP}\Bigr) \sim
(4\pi t)^{-{m/2}}\sum_{n=0}^{\infty} t^{n/2}
A_{n/ 2}(f,P,{\cal B}) .
\eqno (1.6)
$$
The coefficients $A_{n/ 2}(f,P,{\cal B})$ consist of two different parts,
the interior part $C_{n/ 2}(f,P)$ and the boundary part 
$B_{n/ 2}(f,P,{\cal B})$, i.e. 
$$
A_{n/ 2}(f,P,{\cal B})=C_{n/ 2}(f,P)+
B_{n/ 2}(f,P,{\cal B}).
$$
The interior parts $C_{n/ 2}(f,P)$ are 
obtained by integrating some geometric invariants
over $M$ and {\it do not depend} on the boundary conditions.
They are calculated up to $C_4$ in [16]. By contrast,
the boundary parts $B_{n/ 2}(f,P,{\cal B})$ are 
obtained by integrating some geometric invariants
over the boundary $\partial M$ and 
{\it do crucially depend} on the boundary conditions.
It is also well known that, for odd values of $n$, the interior parts 
$C_{n/ 2}(f,P)$ vanish.
In this paper we are mainly interested in the boundary parts.
 
The first few coefficients for 
Dirichlet and Robin boundary conditions with
$$
{\cal B}_{D}=1 \; \; \; \; {\cal B}_{R}=\nabla_{N}+S  
\eqno (1.7)
$$
where $\nabla_N$ is the normal derivative 
and $S$ is a matrix-valued function on the boundary, 
are listed in appendix A.
The crucial point is that the numerical coefficients
multiplying all possible geometric invariants contributing to
the heat-kernel coefficients with the given boundary
conditions are {\it universal constants}, 
independent of the dimension $m$. 

As shown in [10], one can consider for the operator $P$
not only the Dirichlet or Robin boundary conditions, but also
the more general case when the boundary operator takes the form
$$
{\cal B}_A=\nabla_{N}
+{1\over 2}\left(\Gamma^{i}{\widehat \nabla}_{i}
+{\widehat \nabla}_{i}\Gamma^{i}\right) +S 
\eqno (1.8)
$$
where $\Gamma^{i}$ ($i=1,2,\dots,m-1$) and $S$ are 
some matrices which depend on the local
coordinates on the boundary. Symmetry of the operator $P$
is ensured if ${\Gamma^{i}}^{\dagger}=-\Gamma^{i}$ and
$S^{\dagger}=S$. The consideration of boundary operators of
the kind (1.8) is suggested by self-adjointness theory [9, 10],
string theory [17, 18], and BRST invariance [11].
It should be stressed, however, that the results derived in
[10] only hold when the matrices $\Gamma^{i}$ commute with
each other, i.e. $[\Gamma^{i},\Gamma^{j}]=0$
(Abelian case). This non-trivial
point was not realized by the authors of [10], and its proof
is part of recent work by us [19].

Our paper, relying on [5, 10, 14], studies the heat-kernel
asymptotics for the operator $P$ resulting from the
boundary operator (1.8). For this purpose, section 2
describes the problems occurring in the analysis of Euclidean
quantum gravity. The first original results are presented
in section 3, devoted to the investigation of the totally
flat case. Section 4 describes conformal-variation methods
in a way accessible to physics-oriented readers, with 
application to the analysis of the $A_{1}$ coefficient
in the Abelian case.
The structure of $A_{3/ 2}(f,P,{\cal B})$ and 
$A_{2}(f,P,{\cal B})$ is presented in section 5 under three
assumptions, the main one requiring again that the 
matrices $\Gamma^{i}$ commute with each other.
Section 6 is devoted to an original reduction technique
for heat-kernel coefficients. Results and open problems
are described in section 7, and relevant details are given
in the appendix. 

\bigskip\bigskip
\leftline{\mittel 2. Boundary operator in Euclidean quantum gravity}
\vglue0pt
\bigskip
\vglue0pt
As shown in detail in [2, 6, 7, 9, 11], generalized boundary 
conditions involving also tangential derivatives
occur naturally in Euclidean quantum gravity.
The metric perturbations are symmetric rank-two tensors. More
precisely, in geometric language, they are smooth sections of
the vector bundle of such tensors. This bundle has a connection
$
\omega_{e,\, ab}{}^{cd}=-2\Gamma^{(c}{}_{e(a}\delta^{d)}{}_{b)}
$
and a curvature
$
{\Omega}_{ef,\, ab}{}^{cd}=-2R_{ef(a}{}^{(c}\delta_{b)}{}^{d)},
$
and its metric is defined by the equation
$$
H^{ab\; cd}\equiv g^{a(c}g^{d)b}-{1\over 2}g^{ab}g^{cd} .
\eqno (2.1)
$$
The corresponding operator in the covariant de Donder type 
minimal gauge is then of Laplace type (1.1),
with the ``potential" term of the form
$$
E_{ab}{}^{cd}=-2 R_{(a}{}^c{}_{b)}{}^d
+2R_{(a}{}^{(c}\delta_{b)}{}^{d)}
-g_{ab}R^{cd}
-{2\over m-2}R_{ab}g^{cd}
+{2\over m-2}g_{ab}g^{cd}R .
$$
Hereafter, $N^{a}$ is the unit normal to the boundary, 
$e^{i}$ is a local basis for the tangent bundle of the boundary,
$e_i$ is the dual basis on the cotangent bundle,
$g^{ab}e^i{}_{a}e^j{}_{b}=\gamma^{ij}$,
$\gamma^{ij}$ being the induced metric on the boundary, 
$q^{ab}=g^{ab}-N^{a}N^{b}=e_{i}^{a}e_{j}^{b}\gamma^{ij}$;
$\nabla_i\equiv e^a_i\nabla_a$, and
$\nabla_{N} \equiv N^{a}\nabla_{a}$ denotes the normal derivative.
One has
$\nabla_N N=0$, $\nabla_i N=
\nabla_N e_i=K^j_i e_j$,
where $K_{ij}$ is the second fundamental form of a surface 
$\partial M(r)$ obtained by a normal geodesic shifting of 
the boundary to a distance $r$ [5].
Further, we have $\nabla_j e_i=K_{ij}N+\omega^k_{ij}e_k$,
where $\omega^k_{ij}$ is {\it defined} by this equation.
The connection $\widehat\nabla$ of $\partial M(r)$
is {\it defined} by 
$\widehat\nabla_j e_i\equiv \nabla_j e_i
-\omega^k_{ji}e_k$
and is compatible with the metric $\gamma$,
$\widehat\nabla_{i}\gamma_{jk}\equiv 0$, if
the quantities $\omega^k_{ij}$ are the 
Christoffel symbols computed for the metric $\gamma_{ij}$.
Here, as usual, we have included in the 
tangential derivatives $\widehat\nabla$ 
both the standard Levi-Civita connection of the boundary
and the restriction of the bundle connection to the boundary.
In the coordinate basis on the boundary 
one has just $N^{a}=N_{a}=(1,0,\dots,0)$ and
$e^i{}_{a}=(0,\delta^{i}_{a})$. A form of the boundary operator
which leads to boundary conditions completely invariant under
infinitesimal diffeomorphisms of metric perturbations reads [9]
$$
{\cal B}=\mu\Pi+(\II-\Pi)\left(H\nabla_N
+\Gamma^{i}{\widehat \nabla}_{i}+S \right) 
\eqno (2.2)
$$
where $\mu$ is an auxiliary dimensional parameter 
introduced to preserve dimensions, 
$H$ denotes the matrix $H=(H_{ab}{}^{cd})$, 
$\Pi$ is a projector of the form
$$
\Pi_{ab}{}^{cd} \equiv q^{c}{}_{(a}\; q^d{}_{b)} 
\eqno (2.3)
$$
and $\Gamma^{i}$ and $S$ are matrices acting on
symmetric second-rank tensors, of the form [9]
$$
\Gamma^i{}_{ab}{}^{cd}\equiv -N_{a}N_{b}e^{i(c}N^{d)}
+N_{(a} \; e^i{}_{b)}\; N^{c}N^{d} 
\eqno (2.4)
$$
$$
S_{ab}{}^{cd} \equiv 
-N_{a}N_{b}N^{c}N^{d}K
+2N_{(a}e^i{}_{b)}e^{j(c} N^{d)}
\left[K_{ij}+\gamma_{ij} K \right] .
\eqno (2.5)
$$
As is shown in [9], the operator $P$ with the boundary operator
(2.2) is symmetric with respect to the metric (2.1).

We see that the matrices $\Gamma^{i}$ (which are, more precisely,
endomorphism-valued vector fields on the boundary) are antisymmetric
and the matrix $S$ is symmetric. Moreover, one finds
$$
\Gamma^i\Gamma^j=-{1\over 2}\gamma^{ij}N_aN_bN^cN^d
-N_{(a}e^i{}_{b)}e^{j(c}N^{d)}. 
\eqno (2.6)
$$
Therefore, we find that the matrices $\Gamma^{i}$ do not
commute with each other,
$$
[\Gamma^{i},\Gamma^{j}]=-N_{(a} e_{b)}^{i}e^{j(c} N^{d)}
+N_{(a}e_{b)}^{j}e^{i(c} N^{d)}.
\eqno (2.7)
$$
Further, we compute
$$
\Gamma^2=-{1\over 2}(m-1)N_{a}N_{b}N^{c}N^{d}
- N_{(a}q^{(c}{}_{b)}N^{d)} .
\eqno (2.8)
$$
This implies that the matrix $\Gamma^{2}$ commutes with the matrix $S$
$$
\eqalignno{
\Gamma^{2}S & = S\Gamma^{2}
={1\over 2}(m-1)N_{a}N_{b}N^{c}N^{d}K \cr
& - N_{(a} e_{\; \; b)}^{i} e^{j(c} N^{d)}
\left[K_{ij}+\gamma_{ij} K \right] 
&(2.9)\cr}
$$
but does not commute with the matrices $\Gamma^{i}$. 
Indeed, one has
$$
\Gamma^{i}\Gamma^{j}\Gamma^{k}
={1\over 2}N_{a}N_{b}\gamma^{ij}e^{k(c}N^{d)}
- {1\over 2}N_{(a}e^i{}_{b)}\gamma^{jk}N^{c}N^{d}  
\eqno (2.10)
$$
and, hence,
$$
[\Gamma^{2},\Gamma^{i}]
={1\over 2}(m-2)\Bigr[N_{a}N_{b}e^{i(c}N^{d)}
+N_{(a}e^i{}_{b)}N^{c}N^{d}\Bigr]. 
\eqno (2.11)
$$
At a deeper level, as shown in [19, 20], these properties reflect
the lack of strong ellipticity in Euclidean quantum gravity, 
{\it when the boundary operator} (2.2) {\it is used}. 
Without giving any technical
detail, for which we can only refer the reader to the work in
[19, 20], we can say that this means that the trace of the 
heat-kernel diagonal is non-integrable near the boundary. This
property implies, in turn, that the one-loop calculation is 
ill-defined, which is much worst than being just very difficult.
For the time being, one has thus to resort to the analysis of
very simple cases, which is what we begin in section 3.

\bigskip
\bigskip
\leftline{\mittel 3. Heat-kernel asymptotics in a particular case}
\bigskip

When a general structure is investigated, it is appropriate
to show in detail what it implies in the simplest possible
case. This first step may be eventually used to solve 
a more general problem. This can help to determine at least
some class of the terms in the heat-kernel asymptotics.
In this section we evaluate all terms in all the heat-kernel
coefficients that depend only on $E, S$ and $\Gamma^{i}$ (but
not on their derivatives). It is clear that, to calculate these
terms, it is enough to replace the manifold $M$ by a 
{\it flat} manifold $M_{0}={\bf R}_{+} \times \partial M_{0}$,
with a {\it flat} boundary $\partial M_{0}={\bf R}_{m-1}$, and
to assume that all the matrices $E, S$ and $\Gamma^{i}$ are
constant. Note that the manifold $M_{0}$ is non-compact and,
therefore, the trace of the heat kernel does not exist. However,
for a function $f$ with compact support, the trace 
${\rm Tr}_{L^{2}}(f{\rm e}^{-tP})$ does exist and reproduces
correctly the local results. We are actually interested in the
boundary contributions, which are effectively present only in
a narrow strip near the boundary. The difference with respect
to the exact result is always asymptotically small as
$t \rightarrow 0$. In other words, we assume that
$R_{\; bcd}^{a}=\Omega_{ab}=K_{ij}=0$,
$\nabla E={\widehat \nabla}S={\widehat \nabla}\Gamma^{j}=0$.
Moreover, we also assume that the matrices $\Gamma^j$ and $S$ commute
with $E$: $[E,S]=[E,\Gamma^{j}]=0$.
The heat kernel for this problem can be found explicitly.
Denoting the coordinates on the boundary by 
$\hat x^i$ and the normal distance to the boundary by $r$,
and using the Fourier transform in $\hat x^i$, 
one constructs first the parametrix $G(\lambda)$
and then, by the inverse Laplace transform, the heat kernel [19].
Omitting this standard calculation one can present the result in the form
$$
U(t|x,x')=(4\pi t)^{-m/2}{\rm e}^{tE}
\left\{\exp\left[-{1\over 4t}(x-x')^2\right]
+\Omega_B(t|x,x')\right\}
\eqno (3.1)
$$
where
$$
\eqalignno{
\Omega_B(t|x,x')&=\int_{{\bf R}^{m}}{d\zeta_j\,d\omega\over \pi^{m/2}}
%\cr
%&
%\times
\exp\left[-\zeta^2-\omega^2+i\zeta_j{(\hat x^j-\hat x'^j)\over \sqrt t}
+i(\omega+i\varepsilon){(r+r')\over \sqrt t}\right]\cr
&\times
\Biggl\{1-2\bigl(\Gamma^j\zeta_j-i\sqrt t S\bigr)
\Bigl[\omega+i\varepsilon+\Gamma^j\zeta_j-i\sqrt t S\Bigr]^{-1}\Biggr\}
\cr}
$$
where $\varepsilon$ is a positive infinitesimal constant.
Herefrom, by taking the limit $x=x'$ we 
obtain an integral representation of
the heat-kernel diagonal
$$
U(t|x,x)=(4\pi t)^{-m/2}\,
{\rm e}^{tE}\left[1+w(t,r)\right]
\eqno (3.2)
$$
where
$$
\eqalignno{
\; & w(t,r)\equiv \Omega_B(t|x,x)
=\int_{{\bf R}^{m}}{d\zeta_j\,d\omega\over \pi^{m/2}}
\exp\left[-\zeta^2-\omega^2
+2i(\omega+i\varepsilon){r\over \sqrt t}\right]\cr
&\times
\Biggl\{1-2\bigl(\Gamma^j\zeta_j-i\sqrt t S\bigr)
\Bigl[\omega+i\varepsilon+\Gamma^{j}\zeta_{j}
-i\sqrt t S\Bigr]^{-1}\Biggr\}.
&(3.3)\cr}
$$
Note that this result is valid in the 
non-Abelian case when the matrices $\Gamma^j$ 
do not commute with each other and with the matrix $S$. The only condition
is that both $\Gamma^j$ and $S$ should commute with $E$.

Herefrom, by integrating the heat-kernel diagonal 
with a weight function $f$ over $M_0$ 
we obtain the functional trace of the heat kernel
$$ \eqalignno{
\; & {\rm Tr}_{L^{2}}[f{\rm e}^{-tP}]
=(4\pi t)^{-m/2}\int_{M_{0}}{\rm e}^{tE}{\rm Tr}(f) \cr
&+(4\pi t)^{-m/2} \int_{\partial M_{0}}
{\rm Tr} \; {\rm e}^{tE}
\int_{0}^{\infty}dr \; w(t,r)f(r).
&(3.4)\cr}
$$
By taking the limit $t \rightarrow 0$ we recover all 
terms in heat-kernel asymptotics which contain only the matrices
$E, S$ and $\Gamma^{i}$ (without derivatives). It is clear that
the first part, in form of a volume integral, gives only the 
interior contributions and does not depend on $S$ and $\Gamma^{j}$,
whereas the second one gives a purely boundary contribution. 
Thus, by rescaling $r \rightarrow r \sqrt{t}$,
jointly with the Taylor expansion of $f(r\sqrt{t})$ and the integrand in $t$
we find the boundary part of heat-kernel asymptotics.

The first non-trivial coefficient reads [19, 20]
$$ \eqalign{
\; & A_{1/2}(f,P,{\cal B})
=\int_{\partial M}{\rm Tr}\, f\,\,{\sqrt \pi\over 2}
\left\{1 \right. \cr
& \left .+2\int_{{\bf R}^{m-1}}
{d\zeta_j\over \pi^{(m-1)/2}}
{\rm e}^{-\zeta^2}\left[
{\rm e}^{-\Gamma^{i}\Gamma^{j}\zeta_{i}\zeta_{j}}
-1\right]
\right\}. \cr}
$$
This result is valid in the general case of 
{\it non-commuting} matrices $\Gamma^j$.
However, the non-com\-mu\-ta\-ti\-vi\-ty of 
the matrices $\Gamma^i$ makes the analysis
of heat-kernel asymptotics extremely difficult.
Therefore, we consider hereafter the purely 
{\it Abelian} case when the matrices $\Gamma^j$
commute with each other and also with the matrix $S$:
$[\Gamma^i,\Gamma^j]=[\Gamma^j,S]=0$. Then the Gaussian integral 
in the formula for $A_{1/2}$ can be easily evaluated 
(see also eq. (4.15) in the following section),
$$
A_{1/2}(f,P,{\cal B})
=\int_{\partial M}{\rm Tr}\, f\,\,{\sqrt \pi\over 2}\left\{
2(1+\Gamma^2)^{-1/2}
-1\right\}.
$$
This coincides with the result of [10]. We would like to stress
that this is valid only when all the matrices $\Gamma^{j}$ 
commute with each other, but is not valid under the only
assumption that $\Gamma^{2}$ commutes with $\Gamma^{i}$
(for details, see [19, 20]).

Further, in the completely Abelian case (only this 
case will be considered from now on) one finds
$$ 
w(t,r)={\rm e}^{-r^2/t}\left\{1+\int_{0}^{\infty}dz 
(2S\sqrt{t}-z\Gamma^{2})
{\rm e}^{\left[\sqrt{t}zS-{z^{2}\over 4}(1+\Gamma^{2})
-{zr\over \sqrt{t}}\right]}\right\} 
\eqno (3.5)
$$
(which coincides with [10]) and, therefore,
$$
\int_{0}^{\infty}dr \; w(t,r)f(r) \sim
\sum_{k=0}^{\infty}t^{(k+1)/2}\sum_{0\leq n \leq k}
\rho_{n,k}S^{n}f^{(k-n)}
\eqno (3.6)
$$
where $f^{(n)} \equiv \left[{d^{n}\over dr^{n}}f\right]_{r=0}$ 
is the $n$-th normal derivative of $f$,
and $\rho_{n,k}$ are some universal functions 
defined by
$$
\rho_{0,k}(\nu) \equiv {1\over k!}
\biggr[{1\over 2}\Gamma\left({k+1\over 2}\right)
-\Gamma^{2} \alpha_{1,k}(\nu) \biggr]
\eqno (3.7)
$$
and, for $n \geq 1$,
$$
\rho_{n,k}(\nu)
\equiv {1\over n! (k-n)!}
\Bigr[2n \alpha_{n-1,k-n}(\nu)
-\Gamma^{2} \alpha_{n+1,k-n}(\nu)\Bigr]
\eqno (3.8)
$$
with $\nu \equiv (1+\Gamma^{2})/4$.
Here $\alpha_{n,k}$ are the universal integrals
(with $k,n \geq 0$)
$$
\alpha_{n,k}(\nu) \equiv \int_{0}^{\infty}dy
\int_{0}^{\infty}dz \; {\rm e}^{-y^{2}-yz-\nu z^{2}}
\; z^{n} \; y^{k}.
\eqno (3.9)
$$
After making the substitution $y=zu$, these integrals
take the form
$$
\alpha_{n,k}(\nu)=
{1\over 2}\Gamma \Bigr({{n+k}\over 2}
+1 \Bigr) \int_{0}^{\infty}du {u^{k}\over
\left[u^{2}+u+\nu \right]^{(n+k)/2+1}}.
\eqno (3.10)
$$
Interestingly, all the functions $\alpha_{n,k}$ can be expressed
in terms of {\it elementary} functions. The first 2 universal
functions can be computed explicitly,
$$
\alpha_{0,0}(\nu)={1\over 2\sqrt{1-4\nu}}
\Artanh(\sqrt{1-4\nu})
\eqno (3.11)
$$
$$
\alpha_{1,0}(\nu)={\sqrt{\pi}\over 2}
{1\over \sqrt{\nu}(1+2\sqrt{\nu})}
\eqno (3.12)
$$
where
$
\Artanh z \equiv \log {1+z \over 1-z}.
$
The functions $\alpha_{n,k}$ satisfy obvious recursion 
relations
$$
{\partial \over \partial \nu}\alpha_{n,k}(\nu)
=-\alpha_{n+2,k}(\nu).
\eqno (3.13)
$$
Moreover, changing the variables $y \rightarrow \sqrt{\nu}z$
and $z \rightarrow {y/\sqrt{\nu}}$ we find the following
symmetry relation:
$$
\alpha_{n,k}(\nu)=\nu^{(k-n)/2}\alpha_{k,n}(\nu).
\eqno (3.14)
$$
Integrating by parts, one can get many other relations among
the functions $\alpha_{n,k}$. In particular,
$$
\alpha_{1,1}=\left(2\nu {\partial \over \partial \nu}+1
\right) \alpha_{0,0}.
\eqno (3.15)
$$
One can thus generate all the universal functions 
$\alpha_{n,k}$ from $\alpha_{0,0}$ and $\alpha_{1,0}$ only,
and one finds
$$
\alpha_{2m,2l}=(-1)^{m+l}{\partial^{m}\over \partial \nu^{m}}
\left(\nu^{l}{\partial^{l}\over \partial \nu^{l}}
\alpha_{0,0} \right)
\eqno (3.16)
$$
$$
\alpha_{2m,2l+1}=(-1)^{m+l}{\partial^{m}\over \partial \nu^{m}}
\left(\nu^{l+1/2}{\partial^{l}\over \partial \nu^{l}}
\alpha_{1,0}\right)
\eqno (3.17)
$$
$$
\alpha_{2m+1,2l}=(-1)^{m+l}\nu^{l-m-1/2}
{\partial^{l}\over \partial \nu^{l}}
\left(\nu^{m+1/2}{\partial^{m}\over \partial \nu^{m}}
\alpha_{1,0}\right)
\eqno (3.18)
$$
$$
\alpha_{2m+1,2l+1}=(-1)^{m+l}{\partial^{m}\over \partial \nu^{m}}
\left \{ \nu^{l} {\partial^{l}\over \partial \nu^{l}}
\left[\left(2\nu {\partial \over \partial \nu}+1 \right)
\alpha_{0,0}\right] \right \} .
\eqno (3.19)
$$
We list below some of the universal functions $\rho_{n,k}$
explicitly, because they will be used in the following
section. They are (re-expressing them in terms of
$\Gamma^{2}$ for convenience)
$$
\rho_{0,0}={\sqrt\pi\over 2}\left({2\over \sqrt{1+\Gamma^2}}-1\right)
\eqno (3.20)
$$
$$
\rho_{1,1}={2\over (1+\Gamma^{2})}
\eqno (3.21)
$$
$$
\rho_{0,1}={1\over\sqrt{-\Gamma^2}}
\Artanh\sqrt{-\Gamma^2}-{1\over 2}
\eqno (3.22)
$$
$$
\rho_{2,2}={\sqrt \pi\over(1+\Gamma^{2})^{3/2}}
\eqno (3.23)
$$
$$
\rho_{1,2}={\sqrt\pi\over\Gamma^2}
\left(1-{1\over\sqrt{1+\Gamma^2}}\right)
\eqno (3.24)
$$
$$
\rho_{0,2}={\sqrt\pi\over 2}\left(
{\sqrt{1+\Gamma^2}-1\over\Gamma^2}-{1\over 4}\right)
\eqno (3.25)
$$
$$
\rho_{3,3}={4\over 3(1+\Gamma^2)^2}
\eqno (3.26)
$$
$$
\rho_{2,3}={1\over \Gamma^2}\left(
{1\over\sqrt{-\Gamma^2}}\Artanh\sqrt{-\Gamma^2}
-{1\over (1+\Gamma^{2})}\right)
\eqno (3.27)
$$
$$
\rho_{1,3}=-{1\over \Gamma^2}\left(
{1\over\sqrt{-\Gamma^2}}\Artanh\sqrt{-\Gamma^2}-1\right)
\eqno (3.28)
$$
$$
\rho_{0,3}={1\over 4}\left(1+{1\over\Gamma^2}\right)
{1\over\sqrt{-\Gamma^2}}\Artanh\sqrt{-\Gamma^2}
-{1\over 12}\left(1+{3\over\Gamma^2}\right).
\eqno (3.29)
$$
Note that all the universal functions $\rho_{n,k}$ are
analytic at $\Gamma=0$, but have a {\it singularity} for
$\Gamma^{2}=-1$. As shown in [19--21], this occurs because
the boundary-value problem is strongly elliptic only when
$(1+\Gamma^{2})<0$.

In the course of deriving (3.20)--(3.29), we have used also
the following formula for the boundary part of heat-kernel
coefficients (cf (1.6)), which results from (3.4), (3.5) and
the Taylor expansion of ${\rm e}^{tE}$:
$$
B_{{n\over 2}}(f,P,{\cal B})=\int_{\partial M}{\rm Tr}
\sum_{k=0}^{[{(n-1)\over 2}]}
\sum_{j=0}^{n-2k-1}
{1\over k!}\rho_{j,n-2k-1} E^{k}S^{j}
f^{(n-2k-j-1)}.
\eqno (3.30)
$$

\bigskip
\bigskip
\leftline{\mittel 4. Conformal variations and the $A_{1}$ coefficient}
\bigskip

We now describe briefly a method for the calculation of
heat-kernel asymptotics which turns out to be very powerful for the 
boundary conditions without tangential derivatives [14].
This method is based on the behaviour of the heat kernel
under conformal rescalings, together with some simple
exactly solvable cases. 
In this method, one makes a conformal deformation of the whole 
boundary value problem, i.e. the operator $P$ and the boundary 
operator ${\cal B}$, with a deformation parameter $\varepsilon$,
and one studies the behaviour of the heat-kernel coefficients 
$A_{n/ 2}(f,P,{\cal B})$ under this
deformation. This is used to determine 
a set of recurrence relations which reduce the evaluation of
heat-kernel asymptotics to the solution of a system of algebraic
equations for a finite set of numerical coefficients. 
We here present shortly the key elements of this
analysis. They are as follows.

Let $f$ be a smooth function on $M$. Then the conformal 
deformations of the metric and of 
the inward-pointing normal are [14]:
${g}_{ab}(\varepsilon)={\rm e}^{2\varepsilon f}\; g_{ab}$,
${N}^{a}(\varepsilon)={\rm e}^{-\varepsilon f}\; N^{a}$.
The conformal deformation of the potential terms
$E(\varepsilon)$ and $S(\varepsilon)$, 
and of $\Gamma^{i}(\varepsilon)$, 
is chosen in such a way that the operator
$P$ and the boundary operator transform uniformly, i.e.
$P(\varepsilon)={\rm e}^{-2\varepsilon f}P$, 
${\cal B}(\varepsilon)={\rm e}^{-\varepsilon f}{\cal B}$.
One then finds, in particular, that
${\Gamma}^{i}(\varepsilon)={\rm e}^{-\varepsilon f}
\; \Gamma^{i}$. 
The Christoffel symbols of the Levi-Civita connection of $M$
rescale as
$$
{\Gamma}_{\; \; ab}^{c}(\varepsilon)
=\Gamma_{\; \; ab}^{c}
+\varepsilon\left(\delta_{a}^{\; c}\; \nabla_{b}f 
+ \delta_{b}^{\; c} \; \nabla_{a}f
- g_{ab}g^{cd} \nabla_{d}f\right) .
\eqno (4.1)
$$
Further, for the extrinsic-curvature tensor one finds
$$
{K}_{ab}(\varepsilon)=
({\nabla}_{a} \;{N}_{b})(\varepsilon)
={\rm e}^{\varepsilon f}\left(K_{ab}
-\varepsilon N_{a}\nabla_{b}f
-\varepsilon g_{ab}\nabla_Nf\right) 
\eqno (4.2)
$$
which implies
$$
{K}(\varepsilon)={g}^{ab}(\varepsilon) \;{K}_{ab}(\varepsilon)
={\rm e}^{-\varepsilon f}\left[ K 
-\varepsilon(m-1)\nabla_N f\right] .
\eqno (4.3)
$$

As shown in [5, 14], three basic 
conformal-variation formulae hold:
$$
\left[{d\over d \varepsilon}A_{n/ 2}
\Bigr(1,{\rm e}^{-2 \varepsilon f}P, 
{\rm e}^{-\varepsilon f}{\cal B}
\Bigr)\right]_{\varepsilon =0}
=(m-n)A_{n/ 2}(f,P,{\cal B}) 
\eqno (4.4)
$$
$$
\left[{d\over d \varepsilon}A_{n/ 2}
\Bigr(1,P-\varepsilon H,
{\rm e}^{-\varepsilon f}{\cal B}
\Bigr)\right]_{\varepsilon=0}
=A_{{n/ 2}-1}(H,P,{\cal B}) 
\eqno (4.5)
$$
$$
\left[{d\over d \varepsilon}A_{n/ 2}
\Bigr({\rm e}^{-2\varepsilon f}H,
{\rm e}^{-2 \varepsilon f}P,
{\rm e}^{-\varepsilon f}{\cal B}\Bigr)
\right]_{\varepsilon=0}=0 
\eqno (4.6)
$$
where $H$ is some smooth function different from $f$. 
The formulae (4.5) and (4.6) are used to obtain recurrence 
relations for the interior coefficients,
whilst the equation (4.4) can be used to determine recurrence
relations which, jointly with equations resulting from eq. 
(4.6) and from the boundary conditions, may determine
the boundary contributions [5, 14]. More precisely, these 
properties are not enough to determine the whole heat-kernel
asymptotics. For this purpose, one needs some
additional information which can be obtained by
studying some specific cases, e.g. one dimension,
manifolds with very high symmetry, product manifolds,
and so on. In particular, one uses extensively a
reduction Lemma along the lines of section 6.
Hereafter we assume, as already specified 
in sections 1 and 3, that
$\Gamma^{i}$ commutes with $\Gamma^{j}$. Although very restrictive,
this assumption has a physical counterpart consisting of 
complex scalar fields coupled to an electromagnetic field.
Needless to say, this is only a first step towards more
interesting models (e.g. Yang-Mills in curved space with
boundary [19]). 

Before we complete our calculations, it is appropriate to
stress why we have considered the conformal-variation
method. The two main reasons are as follows. 
\vskip 0.3cm
\noindent
(i) The emphasis on conformal-variation techniques has led us
to elucidate the structure of the $A_{3/2}$ and 
$A_{2}$ coefficients in the Abelian case 
(see section 5), which is naturally obtained
within the framework of invariance theory, as shown in [5].
Indeed, the conformal-variation technique cannot lead, by itself,
to the complete calculation of $A_{3/2}$ and $A_{2}$. 
Nevertheless, the results of section 3 and of the present section
will be shown to lead (in section 5) to encouraging progress towards
the evaluation of $A_{3/2}$. Thus, some new insight is definitely
obtained. All this appears quite interesting, since the progress
of knowledge is normally achieved in small steps.
\vskip 0.3cm
\noindent
(ii) There exists already some evidence, 
in the recent literature, that this method
can lead to some advancement of knowledge,
because our equations (4.12) and (4.13) (see below) have been
used in [21] to complete an important investigation of the
$A_{1}$ coefficient.

Note now that the form (1.8) of the boundary operator suggests
immediately that the occurrence of $\Gamma^{i}$ gives rise to
new invariants built from the Riemann curvature tensor, 
the extrinsic-curvature tensor of the 
boundary, $E,S$ and $\Gamma^{i}$ (cf [5]). This simple 
but crucial observation makes it possible to 
start from equation (A1), jointly with
$$
B_{{1/2}}(f,P,{\cal B})=A_{{1/2}}(f,P,{\cal B})
=(4\pi)^{{1/2}}
\int_{\partial M}{\rm Tr}(\gamma f)
\eqno (4.7)
$$
$$
\eqalignno{
B_{1}(f,P,{\cal B})=&
%\cr&
{1\over 6}\int_{\partial M}{\rm Tr}\Biggl\{ \Bigl[f\bigr(
b_{0}K+b_{2}S\bigl) 
+b_{1} f_{;N}\Bigr]
%\cr&
+\Bigr[f(\sigma_{1} 
K_{ij}\Gamma^{i}\Gamma^{j})\Bigr] 
\Biggr\}
&(4.8)\cr}
$$
as our {\it ansatz} for the first 3 coefficients in heat-kernel
asymptotics. Our task is to understand how to use this ansatz,
conformal-variation methods and yet other properties 
(see below) to evaluate $\gamma,b_{0},b_{1},b_{2}$ and
$\sigma_{1}$. The resulting
algorithm could then be used to derive higher-order 
heat-kernel coefficients [5, 14].

First, we use the conformal-variation formula (4.3), 
jointly with the property
$$
\left[{d\over d\varepsilon}(K_{ij}\Gamma^{i}\Gamma^{j})
(\varepsilon)\right]_{\varepsilon=0}
=-f K_{ij} \Gamma^{i} \Gamma^{j}
-f_{;N} \Gamma^{2}.
\eqno (4.9)
$$
The identity (4.4) still
holds, since all our unknown coefficients are not affected by
conformal rescalings (see [14] and below). 
Thus, when $n=2$, eq. (4.4) leads to
$$
\left[{d\over d\varepsilon}A_{1}\Bigr(1,
{\rm e}^{-2 \varepsilon f}P,{\rm e}^{-\varepsilon f}
{\cal B}\Bigr)\right]_{\varepsilon=0}
=(m-2)A_{1}(f,P,{\cal B}) 
\eqno (4.10)
$$
where the right-hand side is obtained from eq. (4.8), whilst the
left-hand side receives contributions from eq. (4.9) and from
the conformal-variation formulae in [5, 14]. One thus finds
an integral over $\partial M$ where the coefficient of
$f_{;N}$ should vanish. This leads to (cf [5, 14])
$$
-b_{0}(m-1)-b_{1}(m-2)+{1\over 2}b_{2}(m-2) 
-(m-4)-\sigma_{1}\Gamma^2=0 .
\eqno (4.11)
$$
Since this algebraic equation should hold for {\it all}
values of $m$, one finds the equations
$$
-b_{0}-b_{1}+{1\over 2}b_{2}-1=0 
\eqno (4.12)
$$
$$
b_{0}+2b_{1}-b_{2}+4-\sigma_{1}\Gamma^{2}=0 .
\eqno (4.13)
$$
It is now clear that conformal-variation methods are, by themselves,
unable to evaluate heat-kernel coefficients, since eqs. (4.12)
and (4.13) only determine $b_{0}$ and $\sigma_{1}$, say, 
in terms of $b_{1}$ and $b_{2}$. 
Thus, one needs some additional information. This can be provided by
considering some simple exactly solvable cases. 
In particular, using the analysis of section 3 we find that
the functions $\gamma$, $b_1$ and $b_2$
are determined by the functions $\rho_{n,k}$ defined
in (3.7)--(3.10):
$$
\gamma={1\over 2\sqrt\pi}\rho_{0,0}\qquad
b_1=6\rho_{0,1}\qquad
b_2=6\rho_{1,1}.
\eqno (4.14)
$$
Thus, under our assumptions, 
we find from the equations (4.12)--(4.14) and (3.20)--(3.22)
the complete result (in agreement with [10],
bearing in mind that $\Gamma^{2}<0$) 
$$
\gamma={1\over 4}\left[{2\over \sqrt{1+\Gamma^{2}}}
-1 \right] 
\eqno (4.15)
$$
$$
b_{1}={6\over \sqrt{-\Gamma^2}} \Artanh(\sqrt{-\Gamma^2})-3
\eqno (4.16)
$$
$$
b_{2}={12\over (1+\Gamma^{2})} 
\eqno (4.17)
$$
$$
b_{0}=-6\biggr[{1\over \sqrt{-\Gamma^2}}
\Artanh(\sqrt{-\Gamma^2})-{1\over (1+\Gamma^{2})} \biggr]
+2
\eqno (4.18)
$$
$$
\sigma_{1}={6\over\Gamma^2}\left({1\over \sqrt{-\Gamma^2}}
\Artanh(\sqrt{-\Gamma^2})-{1\over (1+\Gamma^{2})}
\right) .
\eqno (4.19)
$$
It should be stressed that $\gamma,b_{0},b_{1},b_{2},\sigma_{1}$ are
{\it universal functions}, not affected by the conformal rescalings
and independent of the dimension $m$ (although this is clear from
the procedure leading to eqs. (4.11)--(4.13)). 
One can easily check that, if $\Gamma \rightarrow 0$,
$\gamma, b_{0}, b_{1},
b_{2}$ reduce to the constant values occurring  
for Robin boundary conditions.
All coefficients are analytic in $\Gamma^2$, in a neighbourhood
of $\Gamma=0$.
Equations (4.15)--(4.19) express a deep property, 
because they tell us that, by virtue of the occurrence of
tangential derivatives in the boundary conditions, 
boundary contributions to heat-kernel asymptotics 
in the Abelian case are obtained
by integrating over $\partial M$ linear combinations of the
geometric invariants of the problem, where {\it all}
coefficients of the linear combinations are now universal
functions.

\bigskip\bigskip
\leftline{\mittel 5. Structure of $A_{3/ 2}$ and $A_{2}$
in the Abelian case}
\bigskip

As in section 4, the properties that we are going to derive
hold under the following assumptions:
\vskip 0.3cm
\noindent
(i) The problem is {\it purely Abelian} (this is the most
important assumption), i.e. the matrices $\Gamma^{i}$ commute
with each other: $[\Gamma^{i},\Gamma^{j}]=0$.
\vskip 0.3cm
\noindent
(ii) The matrix $\Gamma^{2} \equiv \gamma_{ij}\Gamma^{i}\Gamma^{j}$,
which automatically commutes with $\Gamma^{j}$ by virtue of (i),
commutes also with the matrix $S$: $[\Gamma^{2},S]=0$.
\vskip 0.3cm
\noindent
(iii) The matrices $\Gamma^{i}$ are covariantly constant with
respect to the (induced) connection on the boundary:
${\widehat \nabla}_{i}\Gamma^{j}=0$.

The latter assumption is made to reduce the 
number of independent invariants.
It is thus easy to see that, as a consequence of the antisymmetry 
and the commutativity of the
matrices $\Gamma^i$, the
following property holds:
$$
\left(\Gamma^{i_1}\cdots\Gamma^{i_{k}}\right)^{\dag}
=(-1)^k\Gamma^{i_{1}}\cdots\Gamma^{i_k} .
$$
Thus, the products of an even number of matrices $\Gamma^i$ are symmetric,
whereas the products of an odd number of 
$\Gamma^i$ are antisymmetric matrices; for any
symmetric matrix $X$ and for any antisymmetric matrix $Y$ one has
$$
{\rm {Tr}}\,\alpha(\Gamma)\Gamma^{i_1}\cdots\Gamma^{i_{2k+1}}X=
{\rm {Tr}}\,\alpha(\Gamma)\Gamma^{i_1}\cdots\Gamma^{i_{2k}}Y=0 
$$
where $\alpha(\Gamma)$ is an arbitrary function of $\Gamma^2$.

As we know, the endomorphisms $E$ and $S$ are symmetric 
matrix-valued functions. Thus, all their covariant derivatives are also
symmetric. Hence
$$
{\rm {Tr}}\,\alpha(\Gamma)\Gamma^{i_1}
\cdots\Gamma^{i_{2k+1}}E_{;k \cdots l}=
{\rm {Tr}}\,\alpha(\Gamma)\Gamma^{i_1}
\cdots\Gamma^{i_{2k+1}}S_{|k \cdots l}=0 .
$$
By contrast, the bundle curvature is antisymmetric jointly with all its
covariant derivatives. Therefore
$$
{\rm {Tr}}\,\alpha(\Gamma)\Gamma^{i_1}\cdots\Gamma^{i_{2k}}
\Omega_{;k \cdots l}=0 .
$$

To obtain the form of $A_{3/ 2}$, one has
to consider, further to eq. (A4), all possible contractions
of the matrices $\Gamma^{i}$ with the geometric objects 
of the form $fK^2$, $fKS$, $f\widehat\nabla K$, 
$f\widehat\nabla S$, $fR$, $f\Omega$ and $f_{;N}K$. 
Similarly, the
general form of $A_{2}$ is obtained by taking all contractions
of $\Gamma^{i}$ terms with geometric contributions 
of the form: i) $fK^3$, $fK^2S$, $fKS^2$, $fRK$, $f\Omega K$, 
$fEK$, $fRS$, $f\Omega S$,
$fK\widehat\nabla K$, $fS\widehat\nabla K$, $fK\widehat\nabla S$,
$fS\widehat\nabla S$, 
$f\widehat\nabla\widehat\nabla K$, 
$f\widehat\nabla\widehat\nabla S$,
$f \nabla R$, $f \nabla \Omega$, $f \nabla E$;
ii) $f_{;N}K^2$, $f_{;N}KS$, $f_{;N}\widehat\nabla K$, 
$f_{;N}\widehat\nabla S$,
$f_{;N}R$, $f_{;N}\Omega$, 
and iii) $f_{;NN}K$ (cf eq. (A5)). By virtue of the properties
described above we are led to write
the following formulae (hereafter, $K \equiv K_{\; i}^{i}$):
$$ 
\eqalignno{
A_{3/ 2}(f,P,{\cal B})&=
{\widetilde A}_{3/ 2}(f,P,{\cal B})
+(4\pi)^{{1/2}}{1\over 384}
\int_{\partial M}{\rm {Tr}}\biggl\{f\biggl[
\sigma_{2}\left(K_{ij}\Gamma^{i}\Gamma^{j}\right)^{2}\cr
&+ \sigma_{3}\left({K_{ij}\Gamma^{i}\Gamma^{j}}\right) K  
+\sigma_{4}K_{il}K_{\; j}^{l} \Gamma^{i} \Gamma^{j}
+ \lambda_{1} K_{ij}\Gamma^{i} \Gamma^{j}S \cr
&
+ \mu_{1}R_{iNjN} \Gamma^{i} \Gamma^{j}
+\mu_{2}R_{\; ilj}^{l} \Gamma^{i}\Gamma^{j} 
+{\widetilde b}_{1}\Omega_{iN}\Gamma^{i}\biggr] \cr
&+ \beta_{1} f_{;N} K_{ij}\Gamma^{i} \Gamma^{j}
\biggl\} 
&(5.1)\cr}
$$
$$
\eqalignno{
A_{2}(f,P,{\cal B})&={\widetilde A}_{2}(f,P,{\cal B})
+{1\over 360}\int_{\partial M}{\rm {Tr}}\biggr\{f\biggr[
\sigma_{5}({K_{ij}\Gamma^{i}\Gamma^{j}})^{3}
+ \sigma_{6} \left({K_{ij}\Gamma^{i}\Gamma^{j}}\right)^{2}K \cr
&+\sigma_{7}{K_{ij}\Gamma^{i}\Gamma^{j}}K_{rl}K^{rl}
+ \sigma_{8}{K_{ij}\Gamma^{i}\Gamma^{j}}K ^{2} 
+\sigma_{9} K_{il} K_{\; j}^{l}\Gamma^{i}\Gamma^{j}K \cr
&+\sigma_{10} K_{il}K_{\; j}^{l}K_{rp}
\Gamma^{i}\Gamma^{j}\Gamma^{r}\Gamma^{p}
+\sigma_{11}K_{il}K^{lr}K_{rj}\Gamma^{i}\Gamma^{j} \cr
&+\sigma_{12}\left(K_{ij}\Gamma^{i}\Gamma^{j}\right)^{2}S
+ \sigma_{13} \left({K_{ij}\Gamma^{i}\Gamma^{j}}S\right) K  
+\sigma_{14}K_{il}K_{\; j}^{l} \Gamma^{i} \Gamma^{j}S \cr
&+\lambda_{2} K_{ij}\Gamma^{i} \Gamma^{j} S^2
+\lambda_{3} K_{ij}\Gamma^{i}S \Gamma^{j} S
+\lambda_{4} EK_{ij}\Gamma^{i} \Gamma^{j} 
+\lambda_{5} R_{iNjN} \Gamma^{i} \Gamma^{j}S \cr 
&+ \lambda_{6}R_{\; ilj}^{l} \Gamma^{i}\Gamma^{j}S 
+{\lambda}_{7} \Omega_{iN}\Gamma^{i}S
+\mu_{3} R_{iNjN}K_{\; l}^{i}\Gamma^{j}\Gamma^{l} \cr 
&+\mu_{4} R_{iNjN} \Gamma^{i}\Gamma^{j}K_{rl}\Gamma^{r}\Gamma^{l} 
+\mu_{5} R_{iNjN} \Gamma^{i} \Gamma^{j}K 
+\mu_{6}  R^l{}_{NlN} K_{ij}\Gamma^{i}\Gamma^{j} \cr
&+\mu_{7} R_{\; \; iqj}^{q} K_{\; l}^{i}\Gamma^{j} \Gamma^{l}
+\mu_{8}R_{\; \; ilj}^{l} \Gamma^{i}\Gamma^{j} K  
+\mu_{9} R^q{}_{iqj} K_{lp}\Gamma^{i} 
\Gamma^{j}\Gamma^{l}\Gamma^{p} \cr 
&+ \mu_{10}R^{ql}{}_{ql} K_{ij}\Gamma^{i} \Gamma^{j}
+\mu_{11}R_{liqj}K^{lq}\Gamma^{i} \Gamma^{j}
+{\widetilde \mu}_{1}K_{\mid ij}\Gamma^{i}\Gamma^{j} \cr
&+ {\widetilde \mu}_{2}K_{ij\mid l}^{\; \; \; \; \; \; \mid j}
\; \Gamma^{i} \Gamma^{l} 
+{\widetilde \mu}_{3}K_{ij \mid rl}
\Gamma^{i}\Gamma^{j}\Gamma^{r}\Gamma^{l}
+\widetilde\lambda_{1}[S_{|i},S]\Gamma^{i} 
+\widetilde\lambda_{2}S_{|ij}\Gamma^{i}\Gamma^{j} \cr
&+{\widetilde \rho}_{1}R_{\; ilj;N}^{l}\Gamma^{i}\Gamma^{j}
+ {\widetilde \rho}_{2}R_{iNjN ; N}\Gamma^{i} \Gamma^{j} 
+ {\widetilde b}_{2}\Omega_{iN;\, N}\Gamma^{i}\biggl] 
\cr
&+ f_{;N}\biggr(
\beta_{2}({K_{ij}\Gamma^{i}\Gamma^{j}})^{2}
+\beta_{3}{K_{ij}\Gamma^{i}\Gamma^{j}}K 
+ \beta_{4}K_{il}K_{\; j}^{l}\Gamma^{i}\Gamma^{j} \cr
&+ \beta_{5}K_{ij}\Gamma^{i} \Gamma^{j}S
+ \beta_{6} R_{iNjN} \Gamma^{i} \Gamma^{j}
+\beta_{7} R_{\; ilj}^{l} \Gamma^{i}\Gamma^{j} \cr 
&+{\beta}_{8} \Omega_{iN}\Gamma^{i}\biggr)
+ f_{;NN}\beta_{9}
{K_{ij}\Gamma^{i}\Gamma^{j}}
\biggr\} .
&(5.2)\cr}
$$
Note that we take here for convenience, 
as our basic geometric invariants, 
the components of the curvature tensors on $M$ but not on $\partial M$: 
$R_{ijkl}$, $R^l{}_{ilj}$, $R_{iNjN}$, $\Omega_{iN}$ and $\Omega_{ij}$. 
Further, with our notation, 
${\widetilde A}_{n/ 2}(f,P,{\cal B})$ is
given by a formula formally identical to the one in 
appendix A, where all the universal constants 
for boundary terms are replaced
by universal functions which only depend on $\Gamma$, and
are denoted by the same symbol. Moreover, 
$\left \{ \sigma_{i}, \lambda_{i}, \mu_{i}, 
{\widetilde b}_{i},
\beta_{i}, {\widetilde \mu}_{i}, {\widetilde \lambda}_{i},
{\widetilde \rho}_{i} \right \}$ are new families of
universal functions. We find it worth stressing that many
further invariants would occur in eqs. (5.1) and (5.2), but
their contributions vanish after taking the matrix traces,
for the reasons described at the beginning of this section.

The analysis of the conformal-variation formulae resulting from
the consideration of eqs. (5.1) and (5.2) provides non-trivial
information. In particular, we focus 
on the $A_{3/2}$ coefficient, and we consider the
algebraic equations among universal functions which are obtained
when all boundary coefficients in eqs. 
(A2)--(A5) are promoted to the role
of universal functions, in agreement with our notation for
${\widetilde A}_{n/ 2}(f,P,{\cal B})$. 
For this purpose, we now
insert eq. (5.1) into eqs. (4.4) and (4.6) when $n=3$. Bearing 
in mind, from the analysis of eq. (4.11), that an algebraic
equation involving $m$, i.e. $A+Bm=0 \; \forall m$, implies
that $A$ and $B$ should vanish separately, we find from here
the following equations for the universal functions 
contributing to $A_{3/2}$:
$$
{1\over 2}c_{0}-2c_{1}+c_{2}-c_{6}=0
\eqno (5.3)
$$
$$
-c_{0}+2c_{1}-c_{2}+3c_{6}+\mu_{1}\Gamma^{2}=0
\eqno (5.4)
$$
$$
-{1\over 2}c_{0}+2c_{1}-2c_{3}-c_{5}+{1\over 2}c_{7}=0
\eqno (5.5)
$$
$$
c_{0}-2c_{1}-c_{2}+2c_{3}-2c_{4}+3c_{5}-c_{7}
-\sigma_{3}\Gamma^{2}-\mu_{2}\Gamma^{2}=0
\eqno (5.6)
$$
$$
-c_{7}+c_{8}-c_{9}=0
\eqno (5.7)
$$
$$
c_{7}-2c_{8}+3c_{9}-\lambda_{1}\Gamma^{2}=0
\eqno (5.8)
$$
$$
-4c_{5}-5c_{6}+{3\over 2}c_{9}-\beta_{1}\Gamma^{2}=0. 
\eqno (5.9)
$$
The conformal-variation formulae leading to eqs. 
(5.3)--(5.9) cannot be presented for length reasons. One can
say, however, that one has to apply repeatedly the formulae
in our section 4 and in the appendix of [14]. 
Following the method in [14], some of these equations are obtained 
by setting to zero in eq. (4.4) the coefficients which multiply
a basis for the integral invariants in eq. (5.1). 

Moreover,
relying on section 3, one can express 10 universal functions 
in terms of the $\rho_{n,k}$ given by (3.20)--(3.29)
(see (5.1) and (A4)):
$$
c_{0}={192\over\sqrt\pi}\rho_{0,0} 
\eqno (5.10)
$$
$$
c_{8}={192\over\sqrt\pi}\rho_{2,2} 
\eqno (5.11)
$$
$$
c_{9}={192\over\sqrt\pi}\rho_{1,2} 
\eqno (5.12)
$$
$$
c_{6}={192\over\sqrt\pi}\rho_{0,2}
\eqno (5.13)
$$
$$
d_{14}=360\rho_{1,1} \; \; 
d_{20}=360\rho_{3,3}
\eqno (5.14)
$$
$$
e_{1}=360\rho_{0,1} \; \;
e_{9}=360\rho_{2,3} \; \;
e_{10}=360\rho_{1,3} \; \;
e_{7}=360\rho_{0,3}.
\eqno (5.15)
$$
In particular, eqs. (A4), (5.1) and (5.3)--(5.13) imply that we
have obtained, in a few steps, 11 equations for the 18 universal
functions contributing to $A_{3/2}$. Thus, one is really
halfway through towards the solution of a non-trivial problem in
heat-kernel asymptotics (although the restrictive assumptions
about commutation of the $\Gamma^{i}$ matrices imply that harder
problems are in sight). The task now remains to use lemmas 
involving product manifolds [14], or different boundary conditions
(cf our (6.5)), to complete the evaluation of $A_{3/2}$. 
In the case of the $A_{2}$ coefficient, the conformal-variation
method leads to a smaller percentage of useful algebraic
equations, and hence we omit here any detailed calculation, 
since it would not be especially enlightening.

\bigskip
\bigskip
\leftline{\mittel 6. Reduction technique for heat-kernel coefficients}
\bigskip

In the calculation of heat-kernel coefficients, {\it any}
additional information is helpful. In the present section we
describe a new method which relates heat-kernel coefficients with
{\it different} boundary conditions, and hence can help to
reduce the calculations of some difficult unknown case to a
well studied problem.

Suppose that $A$ is a first-order differential operator,
with (formal) adjoint $A^{\dagger}$,
from which the second-order elliptic 
self-adjoint operators $P_{1}$ and
$P_{2}$ are built as
$$
P_{1} \equiv A^{\dagger}A \; \;
{\rm {and}} \; \; 
P_{2} \equiv A A^{\dagger} 
\eqno (6.1)
$$
subject to boundary conditions with boundary operators 
${\cal B}_{1}$ and ${\cal B}_{2}$, respectively. In the 
applications, ${\cal B}_{1}$ will be of zeroth-order,
and ${\cal B}_{2}$ of first-order. Let
$$
{\cal B}_{2}=-{\cal B}_{1} A^{\dagger} .
$$
Then the operators $P_{1}$ and $P_{2}$ have 
the same spectrum except for zero-modes, 
and for any smooth function $f$ on $M$ one finds
$$
{\partial \over \partial t}\left[{\rm {Tr}}_{L^2} 
\Bigr(f {\rm e}^{-t P_{1}}\Bigr)
-{\rm {Tr}}_{L^2}\Bigr(f {\rm e}^{-tP_{2}}\Bigr)\right]
={1\over 2}{\rm {Tr}}_{L^2} \Bigr(\Phi(f,A)
{\rm e}^{-tP_{1}}\Bigr) 
\eqno (6.2)
$$
where
$$
\Phi(f,A) \equiv -[A^{\dagger},[A,f]]-[A,f]A^{\dagger}
+[A^{\dagger},f]A .
\eqno (6.3)
$$
In the course of deriving eqs. (6.2) and (6.3), we have used
the antisymmetry of $[ \; , \;]$, the cyclic 
property of the trace jointly with the identity
$$
P_{2} \; {\rm e}^{-t P_{2}}=A \; {\rm e}^{-tP_{1}}
\; A^{\dagger} .
\eqno (6.4)
$$

Suppose that it is possible to choose the operator $A$ in such a
way that $\Phi(f,A)$ reduces to a convenient function (this 
occurs, for example, when $A$ is formally self-adjoint 
or skew-adjoint). Then further
identities among heat-kernel coefficients are derived:
$$
(n-m)\Bigr[A_{n/ 2}(f,P_{1},{\cal B}_1)
-A_{n/ 2}(f,P_{2},{\cal B}_2)\Bigr]
=A_{{n/ 2}-1}(\Phi,P_{1},{\cal B}_1) .
\eqno (6.5)
$$
Equation (6.5) provides a generalization of lemma 3.2 in [14].
If $A$ is a first-order {\it one-dimensional} differential operator
of the form $A={\partial_x}-b$, where $b$ is a real 
function on a closed interval, then
$A^{\dagger}=-{\partial_x}-b$, 
$P_{1}=-{\partial_x^{2}}+b_x+b^2$,
and $P_{2}=-{\partial_x^{2}}-b_x+b^2$, 
with Dirichlet and Robin boundary operators:
${\cal B}_{1}=1$ for $P_{1}$ and 
${\cal B}_{2}=\partial_{x}+b$ for $P_{2}$. 
Moreover, $\Phi$ becomes just a function $\Phi=f_{xx}+2f_{x}b$.
Hence eq. (6.5) leads, in particular, to the lemma 3.2 of [14].

An important problem in heat-kernel asymptotics is, now,  whether 
a generalization of eq. (6.5) exists when the boundary operator
includes tangential derivatives. In the affirmative case, one
could combine the result with the conformal-variation formulae,
to compute the unknown universal functions 
in the Abelian case (cf [14]).

\bigskip\bigskip
\leftline{\mittel 7. Concluding remarks}
\bigskip

The naturally occurring question is what has one learned from
the new formulae derived in sections 3--6. Indeed, the first
non-trivial point is having realized that the analysis of
the totally flat case (section 3) contributes to the evaluation
of a large number of universal functions in the curved case.
This result is not merely technical, but lies at the heart
of further progress in the field (see (3.20)--(3.29) and
(5.10)--(5.15)). Second,
in the Abelian case, the matrix $\Gamma^{i}$
in the boundary operator (1.8) is responsible for a large
number of new geometric invariants which contribute to
heat-kernel asymptotics. Moreover, in the integrands of the
boundary contributions to heat-kernel coefficients, such
invariants are weighted with {\it universal functions}, independent
of conformal rescalings of the background metric.
This leads to a purely algebraic contribution of universal
functions to the conformal-variation formulae. 
Third, the form of the coefficients 
$A_{3/2}(f,P,{\cal B})$ and $A_{2}(f,P,{\cal B})$ has been
obtained for the first time, under the 3 assumptions listed
at the beginning of section 5. They have been found to involve
49 new geometric invariants, with respect to the case when
the matrix $\Gamma^{i}$ does not occur in the boundary operator.
Fourth, new algebraic 
equations among universal functions have been obtained (see
eqs. (4.12), (4.13) and (5.3)--(5.9)). Fifth,
the generalized formula (6.5) has been
derived. This leads to a relation among
heat-kernel coefficients with different boundary conditions.

One has now to complete the calculation of 
the universal functions of section 5.
In the general case, one has to take into account the 
covariant derivatives of the matrices $\Gamma^i$ with respect
to the Levi-Civita (and the bundle) connection of the boundary, 
i.e. terms like $K_{ij}\Gamma^{i \mid j}$, in the
heat-kernel asymptotics. In the case of the $A_{1}$
coefficient, from dimensional arguments it is clear that such terms 
do not contribute. The only possible term 
in $A_{1}$ is $f{\rm {Tr}}{\Gamma^i{}_{|i}}$, 
but this vanishes because of the 
antisymmetry of $\Gamma^i$ (cf [10], where a
generalized DeWitt ansatz for manifolds with boundary was used).
One should also say that covariant derivatives of $\Gamma^{i}$ 
will occur in the heat-kernel asymptotics only polynomially
and order-by-order, not in more general functions as the
$\Gamma^{i}$ themselves do. The situation is analogous to the
appearance of derivatives of the potential in 
Wigner-Kirkwood expansions [22].

Moreover, when the matrices
$\Gamma^{i}$ do not commute among themselves, only the
coefficient $A_{1/2}$ is known so far, following the
recent analysis in [19]. Unfortunately, in that case there
is not even a general ansatz for heat-kernel coefficients. The
solution of this general problem might result from the combined
effect of applying a number of different 
techniques [5, 23], and could
require the development of a few more lemmas (cf section 6).
If these last problems could be solved, one could get a deeper
geometric understanding of one-loop divergences in 
quantum field theory on manifolds with boundary.
The one-loop divergence in four dimensions [2] can indeed be
obtained by setting eventually $f=1$
in the formula for $A_{2}(f,P,{\cal B})$, once that all the
universal functions have been evaluated. A further crucial problem
lies in the analysis of heat-kernel asymptotics for non-minimal
operators [5] subject to mixed boundary conditions 
involving tangential derivatives.
This step has its physical counterpart in the
investigation of quantization schemes in arbitrary gauges [24].
Its consideration is suggested by the results in [19, 20], 
where it is proved that a Laplace type operator on metric
perturbations, jointly with local and gauge-invariant boundary
conditions, is incompatible with strong ellipticity in
one-loop quantum gravity based on Einstein's theory.

All this seems
to add evidence in favour of one-loop quantum gravity having a
deep impact on quantum field theory and 
spectral geometry [2, 5--16, 19--21, 23--25].

\bigskip
\leftline{\mittel Acknowledgements}
\bigskip

We are indebted to Hugh Osborn for drawing our attention
to the work in [10], and to Peter Gilkey and Alexander
Kamenshchik for correspondence and scientific 
collaboration, respectively. Help from Luigi Rosa in the
course of using Mathematica is also gratefully acknowledged. 
Anonymous referees made comments which led to a substantial
improvement of the original manuscript.
The work of IA was supported
by the Alexander von Humboldt Foundation and by the
Deutsche Forschungsgemeinschaft.

\bigskip\bigskip
\leftline{\mittel Appendix {A}}
\bigskip

First, we describe our notation.
Our $\left \{a,b,...,\right \}$ range 
from $0$ through $m-1$ and index a local
orthonormal frame for the tangent bundle of the Riemannian
manifold $M$, whilst
our indices $\left \{i,j,...,\right \}$ 
range from $1$ through $m-1$ and index the 
orthonormal frame for the tangent bundle of the boundary of $M$.
The semicolon $;$ denotes multiple covariant differentiation
with respect to the Levi-Civita (and bundle) connection of $M$, whilst a
stroke $\mid$ denotes covariant differentiation tangentially
with respect to the Levi-Civita 
(and the bundle) connection of the boundary. 
If $N$ is the inward-pointing unit normal to $\partial M$,
the normal derivative of a smooth function $f$ on $M$
is denoted by $f_{;N}\equiv \nabla_Nf$.
The curvature of the bundle connection is denoted by $\Omega_{ab}$.

For Dirichlet or Robin boundary conditions, 
the first heat-kernel coefficients read [14]
$$
A_{0}(f,P,{\cal B})=\int_{M}{\rm {Tr}}\,(f) 
\eqno (A1)
$$
$$
A_{1/ 2}(f,P,{\cal B})=(4\pi)^{{1\over 2}}
\int_{\partial M}{\rm {Tr}}(\gamma f) 
\eqno (A2)
$$
$$
\eqalignno{
A_{1}(f,P,{\cal B})&= {1\over 6} \biggr \{
\int_{M}{\rm {Tr}}\Bigr[f(\alpha_{1}E 
+\alpha_{2}R)\Bigr] \cr
&+ \int_{\partial M} {\rm {Tr}} \Bigr[f\left(b_{0}K
+b_{2}S\right)
+b_{1}f_{;N}\Bigr] \biggr \} 
&(A3)\cr}
$$
$$
\eqalignno{
A_{3/ 2}(f,P,{\cal B})&= (4\pi)^{{1\over 2}}
{1\over 384} \biggr \{ \int_{\partial M} {\rm {Tr}}
\biggr[f \Bigr(c_{0}E +c_{1}R
+c_{2}R^i{}_{NiN}
+c_{3} K ^{2} \cr
&+ c_{4}K_{ij}K^{ij}+c_{7}S K 
+c_{8}S^{2} \Bigr) 
+ f_{;N} \Bigr(c_{5} K 
+c_{9}S \Bigr) \cr
&+c_{6}f_{;NN}\biggr] \biggr \} 
&(A4)\cr}
$$
$$
\eqalignno{
A_{2}(f,P,{\cal B})&= {1\over 360} \biggr \{
\int_{M}f{\rm {Tr}}\Bigr(\alpha_{3}E^{;a}{}_{;a}
+\alpha_{4}R E + \alpha_{5} E^{2}+\alpha_{6}R^{;a}{}_{;a}\cr
&+ \alpha_{7}R^{2}
+\alpha_{8}R_{ab}R^{ab}
+\alpha_{9}R_{abcd}R^{abcd}
+\alpha_{10}\Omega_{ab}\Omega^{ab}\Bigr) \cr
&+ \int_{\partial M}{\rm {Tr}}\biggr[f \Bigr(d_{1}E_{;N}
+d_{2}R_{;N}+d_{3} K ^{\mid i}{}_{\mid i}
+d_{4}K^{ij}{}_{\mid ij} \cr
&+ d_{5}E K 
+d_{6}RK 
+d_{7}R^{i}{}_{NiN} K 
+d_{8}R_{iNjN}K^{ij} \cr
&+ d_{9}R^l{}_{ilj}K^{ij}+d_{10} K^{3}
+d_{11}K_{ij}K^{ij}K 
+d_{12}K^i{}_jK^j{}_{l}K^{l}{}_{i} \cr 
&+ d_{13}\Omega^i{}_{N;i}+d_{14}SE
+d_{15}SR
+d_{16}SR^i{}_{NiN}+d_{17}S K^{2} \cr 
&+ d_{18}SK_{ij}K^{ij}+d_{19}S^{2} K 
+d_{20}S^{3}+d_{21}S^{\mid i}{}_{\mid i}\Bigr) \cr 
&+ f_{;N}\Bigr(e_{1}E +e_{2}R
+e_{3}R^i{}_{NiN}
+e_{4} K ^{2}+e_{5}K_{ij}K^{ij} 
+ e_{8}S K +e_{9}S^{2} \Bigr) \cr 
&+ f_{;NN} \Bigr(e_{6} K 
+e_{10}S \Bigr) + e_{7}f^{;a}{}_{;aN} \biggr]
\biggr \} .
&(A5)\cr}
$$
The numerical constants $\left \{ \alpha_{i}, \gamma,
b_{i},c_{i},d_{i},e_{i} \right \}$
for Dirichlet and Robin boundary conditions (1.7)
are calculated, in particular, in [5, 14] (no confusion
should arise from the use of the symbol $\gamma^{ij}$
for the metric on the boundary in section 2).

\bigskip
\bigskip
\leftline{\mittel References}
\vglue0pt
\bigskip
\vglue0pt
\item{[1]}
Hawking S W 1982 {\it Pont. Acad. Scient. Scr. Varia}
{\bf 48} 563
\item{[2]}
Esposito G, Kamenshchik A Yu and Pollifrone G 1997 
{\it Euclidean Quantum Gravity on Manifolds with Boundary}
({\it Fundamental Theories of Physics} {\bf 85})
(Dordrecht: Kluwer)
\item{[3]}
Grib A A, Mamaev S G and Mostepanenko V M 1994 {\it Vacuum
Quantum Effects in Strong Fields} (St. Petersburg:
Friedmann Laboratory Publishing)
\item{[4]}
Hawking S W 1979 in {\it General Relativity, an Einstein
Centenary Survey}, eds S W Hawking and W Israel
(Cambridge: Cambridge University Press)
\item{[5]}
Gilkey P B 1995 {\it Invariance Theory, the Heat Equation 
and the Atiyah-Singer Index Theorem} (Boca Raton, FL:
Chemical Rubber Company)
\item{[6]}
Barvinsky A O 1987 {\it Phys. Lett.} {\bf 195B} 344
\item{[7]}
Esposito G, Kamenshchik A Yu, Mishakov I V and Pollifrone G
1995 {\it Phys. Rev.} D {\bf 52} 3457
\item{[8]}
Marachevsky V N and Vassilevich D V 1996 {\it Class.
Quantum Grav.} {\bf 13} 645
\item{[9]}
Avramidi I G, Esposito G and Kamenshchik A Yu 1996
{\it Class. Quantum Grav.} {\bf 13} 2361
\item{[10]}
McAvity D M and Osborn H 1991 {\it Class. Quantum Grav.}
{\bf 8} 1445
\item{[11]}
Moss I G and Silva P J 1997 {\it Phys. Rev.} D {\bf 55} 1072
\item{[12]}
Vassilevich D V 1995 {\it J. Math. Phys.} {\bf 36} 3174
\item{[13]}
Hawking S W 1977 {\it Commun. Math. Phys.} {\bf 55} 133
\item{[14]}
Branson T P and Gilkey P B 1990 {\it Commun. Part. Diff. Eq.}
{\bf 15} 245
\item{[15]}
Avramidi I G 1993 {\it Yad. Fiz.} {\bf 56} 245
\item{[16]} 
Avramidi I G 1991 {\it Nucl. Phys.} B {\bf 355} 712
\item{[17]}
Abouelsaood A, Callan C G, Nappi C R and Yost S A 1987
{\it Nucl. Phys.} B {\bf 280} 599
\item{[18]}
Callan C G, Lovelace C, Nappi C R and Yost S A 1987
{\it Nucl. Phys.} B {\bf 288} 525
\item{[19]}
Avramidi I G and Esposito G {\it Gauge Theories on
Manifolds with Boundary} (hep-th/9710048) 
\item{[20]}
Avramidi I G and Esposito G {\it Lack of Strong Ellipticity
in Euclidean Quantum Gravity} (hep-th/9708163)
\item{[21]}
Dowker J S and Kirsten K 1997 {\it Class. Quantum Grav.}
{\bf 14} L169
\item{[22]}
Fujiwara Y, Osborn T A and Wilk S F J 1982 
{\it Phys. Rev.} A {\bf 25} 14
\item{[23]}
McKean H P and Singer I M 1967 {\it J. Diff. Geom.}
{\bf 1} 43
\item{[24]}
Barvinsky A O and Vilkovisky G A 1985 {\it Phys. Rep.}
{\bf 119} 1
\item{[25]}
Kirsten K {\it The $a_{5}$ Coefficient on a Manifold
with Boundary} (hep-th/9708081)

\bye